\documentstyle[preprint,eqsecnum,aps,12pt,psfig]{revtex}
\begin{document}
\newcommand{\bare}[1]{\mathaccent"7017{#1}}
\def\e{\epsilon}
\def\c{{\mathrm c}}
\draft
\title{Five-loop additive renormalization in the $\phi^4$ theory and
amplitude functions of the minimally renormalized
specific heat in three dimensions}
\author{S.~A.~Larin\thanks{Permanent address: Institute for Nuclear Research
of the Russian Academy of Science, 60$^{\mathrm th}$ October Anniversary
Prospect 7-A, Moscow 117312, Russia}, 
M.~M\"onnigmann, M.~Str\"osser, and V.~Dohm}
\address{Institut f\"ur Theoretische Physik, Technische Hochschule Aachen,
D--52056 Aachen, Germany}
\maketitle
\begin{abstract}
We present an analytic five-loop calculation for the additive
renormalization constant $A(u,\epsilon)$ and the associated 
renormalization-group function $B(u)$ of the specific heat of the O$(n)$ 
symmetric $\phi^4$ theory within the minimal subtraction scheme. 
We show that this calculation does not require new five-loop 
integrations but
can be performed on the basis of the previous five-loop 
calculation of the four-point 
vertex function combined with an appropriate identification of symmetry
factors of vacuum diagrams.
We also determine the amplitude function $F_+(u)$ of the specific heat in 
three dimensions for $n=1,2,3$  
above $T_\c$ and $F_-(u)$ for $n=1$ below $T_\c$ up 
to five-loop order, without using the $\e=4-d$ expansion.
Accurate results are obtained from Borel resummations of $B(u)$ for
$n=1,2,3$ and of the amplitude functions for $n=1$.
Previous conjectures regarding the smallness of the resummed higher-order
contributions are confirmed.
Combining our results for $B(u)$ and $F_+(u)$ for $n=1$, 2, 3 with those
of a recent three-loop calculation of $F_-(u)$ for general $n$ in $d=3$
dimensions we calculate
Borel resummed universal amplitude ratios $A^+/A^-$ for $n=1$, 2, 3. 
Our result for $A^+/A^-=1.056\pm0.004$ for $n=2$ is
significantly more accurate than the previous result obtained from the
$\e$ expansion up to $O(\e^2)$ and agrees well
with the high-precision experimental result $A^+/A^-=1.054\pm0.001$ for $^4$He
near the superfluid transition obtained from a recent experiment in space.
\end{abstract}
\vspace{5mm}
\pacs{05.70.Jk, 11.10.Gh, 64.60.Ak, 67.40.Kh}
\newpage
\section{Introduction}
\label{sec:intro}

One of the fundamental achievements of the renormalization-group (RG)
theory of critical phenomena is the identification of universality classes
in terms of the dimensionality $d$ of the system and the number $n$
of components of the order parameter \cite{PHA}.
Specifically, RG theory predicts that the critical exponents, 
certain amplitude ratios and scaling functions are universal quantities 
that do not depend, e.g., on the strength of the interaction or on 
thermodynamic 
variables (such as the pressure). The superfluid transition of $^4$He
belongs to the $d=3$, $n=2$ universality class and provides a {unique}
opportunity for an experimental test of the universality prediction by
means of measurements of the critical behavior at various pressures
$P$ along the $\lambda$-line $T_\lambda (P)$. Early tests have been
performed by Ahlers and collaborators and consistency with the universality
prediction was found within the experimental resolution \cite{A1}.
At a significantly higher level of accuracy, the superfluid density and the 
specific heat (or, equivalently, thermal expansion coefficient)
above and below $T_\lambda (P)$ are planned to be measured in the
Superfluid Universality Experiment (SUE) \cite{LDID} under microgravity
conditions or at reduced gravity in the low-gravity simulator \cite{LLI}.
As demonstrated recently \cite{LS}, this would allow to perform 
measurements up to $|t|\simeq 10^{-9}$ in the reduced temperature 
$t = (T - T_\lambda (P))/T_\lambda(P) $.

On the theoretical side, the corresponding challenge is to calculate
as accurately as possible the properties of the O($n$) symmetric
$\phi^4$ model in three dimensions. To extract the leading critical 
exponents from the experimental data and to demonstrate their universality
at a highly quantitative level requires detailed knowledge on the 
ingredients of a nonlinear RG analysis \cite{D1}. They include not only
the well-known RG exponent functions of the $\phi^4$ model whose
fixed point values determine the critical exponents but also the less
well-known amplitude functions \cite{D2,SD1,SD2,KSD,HD,BSD} which contain
the information about universal ratios of leading and subleading amplitudes
\cite{PHA}.

The existing theoretical predictions on the critical exponents \cite{GZ}
within the
minimal subtraction scheme \cite{HV1,A} are based
on field-theoretic calculations to five-loop order \cite{VKT,CGLT,GLTC,KS}
and Borel resummation. By contrast, the present theoretical 
knowledge of the amplitude ratios for $n>1$ below $T_\c$ is based only 
on low-order (mainly 1- and 2-loop) calculations which imply an 
uncertainty at the level of at least 10--30\% \cite{PHA}. It has therefore 
been proposed \cite{D3} to significantly reduce this uncertainty by
performing new higher-order field-theoretic calculations and Borel
resummations of various amplitude functions in three dimensions. 

Both conceptual and computational steps towards this goal have already 
been performed. The conceptual progress includes the demonstration that the
$d = 3$ field theory suggested by Parisi \cite{Parisi} can well be
realized within the minimal subtraction scheme at $d=3$ \cite{D2,SD1,SD2} 
by incorporating Symanzik's non-vanishing mass shift \cite{S} and that
spurious Goldstone singularities for $n>1$ below $T_\c$
can well be treated within this approach \cite{BSD} by using an appropriately
defined pseudo-correlation length \cite{SD2}. The computational steps
include the determination of the amplitude functions 
$F_+(u)$ and $F_-(u)$ of the specific
heat in three dimensions above $T_\c$ for $n=1,2,3 $ \cite{KSD} and 
below $T_\c$ for $n=1$ \cite{HD}, respectively, up to five-loop order, 
and their Borel resummation. These calculations, however,
were not yet complete because of an approximation regarding the additive
renormalization $A(u,\varepsilon)$ of the specific heat and the associated
RG function $B(u)$. Due to the lack of knowledge in the literature about 
higher-order terms, $A(u,\varepsilon)$ and $B(u)$ were approximated by their
two-loop expressions. Although the good agreement between low-order $d=3$
perturbation results  
\cite{D2,D4,SD3} and accurate experiments \cite{A1,GA,A2} provided some 
indication for the smallness of
the effect of the higher-order terms of $B(u)$, no reliable estimate could
be given for  the remaining uncertainty of
$F_\pm(u)$ which could well be of relevance at the level of accuracy 
anticipated in future experiments \cite{LDID}. Furthermore we recall that any
inaccuracy of $B(u)$ enters not only the formulas \cite{SD2} for several 
universal amplitude ratios but also the formulas needed to determine the
effective coupling $u(l)$ from the specific heat \cite{D2,D4,SD3}.

It is the purpose of the present paper to provide the missing information
on the higher-order terms of $A(u,\varepsilon)$ and $B(u)$ by means of a 
new five-loop calculation. We shall show that the analytic calculation of 
$A(u,\varepsilon)$  and $B(u)$ can be directly related 
to the previous calculations
\cite{VKT,CGLT,GLTC,KS} of the four-point vertex function.
This provides the crucial simplification that no new evaluations of 
three-, four- and five-loop integrals are necessary but that only a
new determination of symmetry and O($n$) group factors
of vacuum diagrams is sufficient. 

Using the 
five-loop expression of $B(u)$ we are in the position to determine the 
correct higher-order terms of the minimally renormalized
amplitude functions $F_+(u)$ for $n=1,2,3$ and $F_-(u)$ for $n=1$ in three
dimensions on the basis of previous work \cite{NMB,BG,BB,BBM} 
where a different renormalization scheme was used.  The
new coefficients of the higher-order terms of $F_+(u)$  turn out to differ
considerably from the previous approximate coefficients \cite{KSD} whereas
the coefficients of $F_-(u)$ are only weakly affected by the new higher-order
terms of $B(u)$.

We also perform new Borel resummations of $B(u)$ for $n = 1,2,3$ as well
as of $F_-(u)$ and of $F_-(u)-F_+(u)$ for $n=1$. It turns out
that the result of the Borel resummation for $B(u)$ including the new terms
up to five-loop order differs from the two-loop result $B(u)=n/2+O(u^2)$ 
by only less than 1\% at the fixed point. For the amplitude
functions, our new Borel resummation results differ from the previous ones
\cite{KSD,HD} by about 1\% for $F_-$ and by less than 0.1\%
for $F_--F_+$ at the fixed point. This is a nontrivial and important 
confirmation of the previous conjectures about the smallness of resummed
higher-order contributions \cite{SD2,KSD,HD}.

As a first application, we calculate the universal ratios $A^+/A^-$ and
$a^+_c/a^-_c$ of the leading and subleading amplitudes of the specific
heat above and below $T_\c$ for $n=1$ and $d=3$.
In addition we calculate Borel resummation values of $A^+/A^-$ for $n=2$ and 3
by combining our present results for $B(u)$ and $F_+(u)$ for $n=2$ and 3 with
those of a recent three-loop calculation of $F_-(u)$ for general $n$
\cite{SLD}. All of our calculations are performed at $d=3$ dimensions, without
using the $\e$ expansion.
Our result $A^+/A^-=1.056\pm0.004$ for $n=2$ is more accurate than
the previous result \cite{B} $1.0294\pm0.0134$ obtained from the $\e=4-d$
expansion up to $O(\e^2)$ and agrees well with the high-precision experimental
result \cite{LS} $1.054\pm0.001$ for $^4$He near the superfluid transition
obtained from a recent experiment in space.

\section{Additive renormalization of the specific heat}
\label{sec:add_ren}

The O($n$) symmetric $\phi^4$ model is defined by the usual
Landau-Ginzburg-Wilson functional
\begin{equation}
\label{eq:LGW}
{\cal H}\{\vec\phi_0({\bf x})\} = \int_V \mbox{d}^{d}x
\left(\frac{1}{2}r_0\phi_0^2 + \frac{1}{2}\sum_{i}
(\nabla\phi_{0i})^2
+u_0(\phi_0^2)^2 -\vec{h}_0\cdot\vec\phi_0\right)
\end{equation}
for the $n$-component field
$\vec\phi_0({\bf x})=(\phi_{01}({\bf x}),\ldots,\phi_{0n}({\bf x}))$
where
\begin{equation}
\label{eq:r0}
r_0=r_{0\c}+a_0t\,, \quad t=(T-T_\c)/T_\c\,,
\end{equation}
and $\vec{h}_0=(h_0,0,\ldots,0)$.
The Gibbs free energy per unit volume (divided by $k_BT$) is
\begin{equation}
F_0(r_0,u_0,h_0) = -V^{-1} \ln\int\! {\cal D}
\vec{\phi}_0\,\exp(-{\cal H}) \,.
\label{eq:gibbs}
\end{equation}
We shall consider the bulk limit $V\to\infty$. We are interested in the
specific heat $\bare{C}^\pm$ per unit volume at vanishing external field
$h_0=0$ (divided by Boltzmann's constant $k_B$) where $\pm$ refers
to $T>T_\c$ and $T<T_\c$, respectively. Near $T_\c$, $\bare{C}^\pm$ is
determined by \cite{SD2}
\begin{equation}
\label{eq:C+-}
\bare{C}^\pm = C_B - T_\c^2 \frac{\partial^2}{\partial T^2} F_0 (r_0,u_0,0)
= C_B - a_0^2 \frac{\partial^2}{\partial r_0^2} F_0(r_0,u_0,0)
\end{equation}
where $C_B$ is an analytic background term. Alternatively the
Helmholtz free energy per unit volume $\Gamma_0(r_0,u_0,M_0) 
= F(r_0,u_0,h_0) + h_0M_0$ with $M_0\equiv\langle\phi_{01}\rangle$ 
determines $\bare{C}^\pm$ in the $h_0\to0$ limit according to
\begin{equation}
\label{eq:C+-Gamma}
\bare{C}^\pm = C_B - a_0^2 \frac{\mathrm d^2}{{\mathrm d}r_0^2} 
\Gamma_0\left( r_0,u_0,M_0(r_0,u_0) \right) \,.
\end{equation}
The perturbative expression for $\Gamma_0(r_0,u_0,M_0)$
is obtained from the negative sum of all one-particle irreducible (1 PI)
vacuum diagrams \cite{A}. The perturbative expression for $\bare{C}^\pm$
is then determined by the vertex functions
$\bare{\Gamma}_\pm^{(2,0)}=\mbox{d}^2\Gamma_0/\mbox{d}r_0^2$
which we consider as functions of appropriately defined correlation lengths
$\xi_+$ and $\xi_-$ above and below $T_\c$ \cite{SD2,BSD},
\begin{equation}
\label{eq:C(xi)}
\bare{C}^\pm = C_B - a_0^2 \bare{\Gamma}_\pm^{(2,0)} (\xi_\pm,u_0,d)\,.
\end{equation}
A description of the critical behavior requires to turn to the renormalized
vertex functions
\begin{equation}
\label{eq:C_ren}
\Gamma_\pm^{(2,0)}(\xi_\pm,u,\mu,d) = Z_r^2 \bare{\Gamma}_\pm^{(2,0)}
(\xi_\pm, \mu^\e Z_u Z_\phi^{-2} A_d^{-1} u,d) - \frac{1}{4} 
\mu^{-\e} A_d A(u,\e)\,.
\end{equation}
We work at infinite cutoff using the prescriptions of dimensional 
regularization and minimal subtraction at fixed dimension $2<d<4$
without employing the $\e=4-d$ expansion \cite{D2,SD1,SD2}. 
The $Z$-factors are introduced as
\begin{equation}
\label{eq:renorm}
r = Z_r^{-1}(r_0-r_{0\c}), \quad
u = \mu^{-\epsilon}A_{d}Z_u^{-1}Z_\phi^2u_0, \quad
\vec\phi = Z_\phi^{-1/2}\vec\phi_0 \label{eq:renormalization} 
\end{equation}
where the geometric factor
\begin{equation}
A_d=\Gamma (3-d/2)\, 2^{2-d} \pi^{-d/2} (d-2)^{-1}
\label{eq:Ad}
\end{equation}
becomes $A_3=(4\pi)^{-1}$ for $d=3$ and $A_4=(8\pi^2)^{-1}$ for $d=4$. 
These $Z$-factors $Z_i(u,\e)$ and the associated
field-theoretic functions \cite{SD1}
\begin{eqnarray}
\label{eq:zeta_r}
\zeta_r(u) &=& \mu \left. \partial_\mu \ln Z_r(u,\e)^{-1} \right|_0 \,, \\
\label{eq:zeta_phi}
\zeta_\phi(u) &=& \mu\left. \partial_\mu \ln Z_\phi(u,\e)^{-1} \right|_0 \,, \\
\label{eq:beta_u}
\beta_u(u,\e) &=& -\e u + \tilde{\beta}(u) = u
\left[ -\e + \mu \left. \partial_\mu (Z_u^{-1}Z_\phi^2) \right|_0\, \right]\,,
\end{eqnarray}
are known up to five-loop order \cite{CGLT,GLTC,KS}.\\

The main quantity of interest in the present paper is the renormalization
constant $A(u,\e)$ in Eq.~(\ref{eq:C_ren}) which absorbs the additive poles
of both $\bare{\Gamma}_+^{(2,0)}$ and $\bare{\Gamma}_-^{(2,0)}$.
Previously \cite{KSD,HD} $A(u,\e)$ was employed only in its two-loop form 
\cite{D2}
\begin{equation}
A(u,\epsilon)=-2n\frac{1}{\epsilon}-8n(n+2)\frac{u}{\epsilon^2}
+O(u^2)\,. \label{eq:A_twoloop}
\end{equation}
Here we report on a calculation of $A(u,\e)$ up to five-loop order.
We would like to stress that this calculation does not require new five-loop 
integrations but
can be performed on the basis of the previous five-loop 
calculation \cite{CGLT,GLTC,KS} of the four-point 
vertex function combined with an appropriate identification of symmetry and
O($n$) 
group factors of vacuum diagrams which are shown in Fig.~\ref{figure:vacuum}.
Their negative sum determines the Helmholtz 
free energy $\Gamma_0$ up to five-loop order.
For the present purpose of determining the pole terms at $d=4$ it suffices
to consider only the case $r_0>0$ and $M_0=0$ where only four-point vertices
exist.
The diagrams are labeled $(1)$ in one-loop order, $(2)$ in two-loop order, 
$(3)$, $(4)$ in three-loop order, $(5)$--$(8)$ in
four-loop order and $(9)$--$(18)$ in five-loop order. The analytic 
expression of an $m$-loop diagram $(i)$ is given by the product of
the coupling $(-u_0)^{m-1}$, the symmetry factor $S^{(i)}$, the O($n$) 
group factor $G^{(i)}(n)$ and the 
momentum integral expression $I^{(i)}(r_0,\e)$.
Thus the structure of the diagrammatic 
expression of a typical diagram, e.g., $(16)$ is
\begin{eqnarray}
\raisebox{-6mm}{\psfig{figure=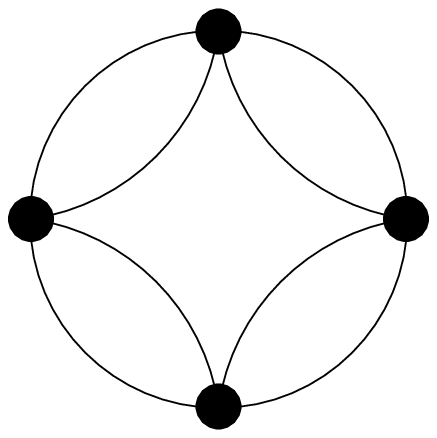,width=15mm}}
\quad &=&\quad S^{(16)} (-u_0)^4\, 
G^{(16)}\, I^{(16)}(r_0,\e) \\
&=& \label{eq:sample_F}
\quad 2592\, (-u_0)^4 \, \frac{n^4 +8n^3 +32n^2 +40n}{81} \nonumber\\
&& \mbox{} \times \int_{p_1}\!\int_{p_2}\!\int_{p_3}\!\int_{p_4}\!\int_{p_5}
G_1\cdot G_2\cdot G_{1+2-3}\cdot G_3\cdot G_{1+2-4}\cdot G_4\cdot G_{1+2-5}
\cdot G_5
\end{eqnarray}
with $\int_p \equiv (2\pi)^{-d} \int {\mathrm d}^dp$ and the propagators
$G_{i\pm j} \equiv (r_0 + |{\mathbf p}_i \pm {\mathbf p}_j |^2)^{-1}$.
The symmetry and group factors are listed in Table \ref{table:symmetry}.

To calculate the additive renormalization constant $A(u,\e)$ one needs to
calculate those ultraviolet $d=4$ pole terms of the diagrams contributing to
$\Gamma_{0+}^{(2,0)}$ that are left after subtraction of
subdivergences.
One obtains $\Gamma_{0+}^{(2,0)}$ by taking two derivatives of $\Gamma_0$
with respect to $r_0$.
The analytic calculation of the poles of the diagrams
for $\Gamma_{0+}^{(2,0)}$
is identical to that carried out previously \cite{CGLT,GLTC,KS} 
for the four-point vertex function $\Gamma_0^{(0,4)}$.
To see this, one should take into account that in
the minimal subtraction scheme \cite{HV1} the ultraviolet pole terms 
specified above do not depend on $r_0$. Then by using
the method of infrared rearrangement \cite{V} one can nullify $r_0$ 
and introduce for each diagram a new fictitious external momentum to
regularize infrared divergences.
Then one can see that only a particular subset of those diagrams 
of $\Gamma_0^{(0,4)}$
are relevant in the present context, namely those where the four external legs
are connected to each diagram through only two four-point vertices (rather
than three four-point vertices or four four-point vertices).
The details of the calculation are presented in Appendix A. 

The result reads
\begin{equation}
\label{eq:A(u,e)}
A(u,\e) = \sum_{m=1}^5 A^{(m)}(u,\e) + O(u^5)
\end{equation}
where $A^{(m)}$ denotes the contribution of $m$-loop order,
\begin{eqnarray}
\label{eq:A3(u,e)}
A^{(3)}(u,\e) &=& - \frac{4}{3}\, n(n+2) \left[
\frac{3}{\e} -\frac{40}{\e^2} +\frac{24(n+4)}{\e^3} \right] u^2 \,, \\
\label{eq:A4(u,e)}
A^{(4)}(u,\e) &=& - \frac{8}{3}\, n(n+2) \left[ 
\frac{(n+8)(12\zeta(3) - 25)}{\e} +\frac{96n+696}{\e^2} \right. \nonumber\\
&& \left. - \frac{248n+1024}{\e^3} + \frac{48(n+4)(n+5)}{\e^4} 
   \right] u^3\,, \\
\label{eq:A5(u,e)}
A^{(5)}(u,\e) &=& - \frac{2}{15}\, n(n+2) \left[
        \frac{768(n+4)(n+5)(5n+28)}{\epsilon^5}
       -\frac{128(293n^2+2624n+5840)}{\epsilon^4} \right. \nonumber\\
       &&  \mbox{} 
       +\frac{9216\zeta(3)(5n+22)+32(519n^2+8462n+25048)}{\epsilon^3} 
       \nonumber\\
       &&  \mbox{} 
       -\frac{192\zeta(3)(7n^2-28n+48) +4608\zeta(4)(5n+22)
       +64(31n^2+2354n+9306)}{\epsilon^2} \nonumber\\
       &&  \mbox{} 
       +\Big( 48\zeta(3)(3n^2-382n-1700) +288\zeta(4)(4n^2+39n+146) \nonumber\\
       &&  \mbox{} \left. 
               -3072\zeta(5)(5n+22) -3(319n^2-13968n-64864) \Big)
        \frac{1}{\epsilon}
                                                  \right] u^4\,, \hspace*{-2mm}
\end{eqnarray}
where $\zeta(s)=\sum_{j=1}^\infty j^{-s}$ is the Riemann zeta function
with $\zeta(3)=1.20205690$, $\zeta(4)=\pi^4/90$ and $\zeta(5)=1.03692776$.
Most important is the $d$-independent RG function $B(u)$
which is determined by \cite{SD2}
\begin{equation}
\label{eq:def:B(u)}
4 B(u) = \left[ 2\zeta_r(u) - \e \right] A(u,\e) + \beta_u(u,\e) \frac{\partial
A(u,\e)}{\partial u} \,.
\end{equation}
Using $A(u,\e)$ of Eqs.~(\ref{eq:A(u,e)})--(\ref{eq:A5(u,e)}) and the
perturbative expressions for $\zeta_r$ and $\beta_u$ of Refs.\
\cite{CGLT,GLTC,KS} we
find
\begin{eqnarray}
\label{eq:B(u)}
B(u) &=& \frac{n}{2} +3n(n+2)u^2 
-\frac{8}{3}n(n+2)(n+8)(25-12\zeta(3))u^3 \nonumber\\
&& \mbox{} +\frac{1}{2}n(n+2) \left[ 16\zeta(3) (3n^2-382n-1700) 
-1024\zeta(5) (5n+22) \right. \nonumber\\ 
&& \mbox{} \left. +96\zeta(4) (4n^2+39n+146) -319n^2+13968n+64864 \right]\!
u^4 \nonumber\\ 
&& \mbox{} + O(u^5).
\end{eqnarray}
The terms of $O(u^2)$ and $O(u^3)$ agree with those of Ref.~\cite{Kas}.
In Table \ref{table1} the coefficients $c_{{}_{Bm}}$ of the power series
\begin{equation}
\label{eq:power_B(u)}
B(u) = \sum_{m=0}^\infty c_{{}_{Bm}} u^m
\end{equation}
are given for $n=0,1,2,3$ up to $m=4$ corresponding to five-loop order. 
Table \ref{table1} also contains the coefficients $f_i^{(m)}$ of the power
series of the functions
\begin{equation}
\label{eq:f_i(u)}
f_i(u) = \sum_{m=1}^\infty f_i^{(m)} u^m
\end{equation}
where $f_i(u)$ denotes the functions $\tilde{\beta}(u)$, $\zeta_r(u)$ and
$\zeta_\phi(u)$ for $i=1,2,3$, respectively.
These coefficients are taken from Refs.\ \cite{CGLT,GLTC,KS}.
Up to four-loop order they agree with those in Table 1 of Ref.\ \cite{SD1}.
(Note that $f_i^{(k)}$ in the table caption of Ref.\ \cite{SD1} should read
$f_i^{(k)} \times 10^{-4}$.)
The five-loop coefficients $f_1^{(6)}$, $f_2^{(5)}$ and $f_3^{(5)}$
differ from those in Table 1 of Ref.~\cite{SD1} 
according to the corrections
in five-loop order in Ref.~\cite{KS}.

In Fig.\ \ref{figure:B} the partial sums 
of $B(u)$ from two- to five-loop order are shown
for the example $n=2$. As expected, the contributions for $m\geq2$ have
alternating signs and increase considerably 
in magnitude. Clearly a resummation of $B(u)$
is necessary similar to that for $\zeta_r(u)$, $\zeta_\phi(u)$ and 
$\tilde{\beta}(u)$ performed previously \cite{SD1}.

First we reexamine the fixed point values $u^\star$, $\beta_u(u^\star,1)=0$,
for $n = 1,2,3$ obtained in Refs.\ \cite{SD1,SD3}
by means of Borel resummation
on the basis of previous five-loop results \cite{CGLT,GLTC}
and in Ref.~\cite{VKT} on the basis of four-loop results.
Here we employ the corrected five-loop coefficients for
the $\e$ expansion of the fixed point value which
we have derived from Eq.~(8) of Ref.~\cite{KS}.
Employing the standard Borel resummation method \cite{SD1,GZJ} we
have obtained the fixed point values in three dimensions
\begin{eqnarray}
\label{eq:u1}
u^\star &=& 0.0404 \quad \pm 0.0003 \qquad \mbox{for } n = 1\,, \\
\label{eq:u2}
u^\star &=& 0.0362 \quad \pm 0.0002 \qquad \mbox{for } n = 2\,, \\
\label{eq:u3}
u^\star &=& 0.0327 \quad \pm 0.0001 \qquad \mbox{for } n = 3\,. 
\end{eqnarray}
The corresponding resummation parameters $\alpha$ and $b=5.5+n/2$ 
\cite{GZJ} are 
\begin{eqnarray}
\label{eq:par1}
2.22 \leq\alpha \leq 3.41\,, & \quad b=6.0 & \qquad \mbox{for }n=1\,, \\
\label{eq:par2}
2.45 \leq\alpha \leq 3.43\,, & \quad b=6.5 & \qquad \mbox{for }n=2\,, \\
\label{eq:par3}
2.71 \leq\alpha \leq 3.43\,, & \quad b=7.0 & \qquad \mbox{for }n=3\,.
\end{eqnarray}
The previous fixed point values \cite{SD1,VKT,SD3} are consistent
with Eqs.~(\ref{eq:u1})--(\ref{eq:u3}) 
within the previous error bars. The present
error bars are smaller than the previous ones \cite{SD1,VKT,SD3}. 
(The range of $\alpha$ determines our error bars, as described further
below.)

We have performed Borel resummations of $B(u)$ at the fixed point $u^\star$
for the cases $n=1$, 2, 3. In addition, for the important case $n=2$
(superfluid $^4$He), we have determined the Borel resummed function
$B(u)$ at various values of $u$.
The results are given in Eq.~(\ref{eq:B1})--(\ref{eq:B3}) and in Figs.\ 
\ref{figure:B} and \ref{figure:B_detail}.

A description of the Borel resummation method \cite{GZJ} 
for the present purpose
has been given in Section 5 of Ref.\ 
\cite{SD1}. (In Eq.~(5.10) of Ref.\ \cite{SD1} 
$a_{jk}$ should read $a_{j-m, k}$.)
In the present work, however, 
we use a different way of determining the parameters 
$\alpha$ and $b$ of the summations. 
This implies a different determination of the error bars.
 
For $B(u)$ the value of the parameter $b$ is not known from an 
analysis of the large-order behavior (see Eq.~(5.6) of Ref.~\cite{SD1} 
and references therein). Here we fix both $b$ and $\alpha$ by requiring 
fastest convergence of the series of the partial Borel sums $S^{(L)} =
S^{(L)}(u,\alpha,b)$ for $B(u)$ defined in Eq.~(5.12) of Ref.~\cite{SD1}
(here $L$ corresponds to $(L+1)$-loop order).
To do so we look for the minima of 
$\Delta^{(4)}$ and $\Delta^{(3)}$ with regard to variations of both $b$ and
$\alpha$ where
\begin{equation}
\label{eq:Delta_l}
\Delta^{(L)}(u,\alpha, b)= \left| \frac{S^{(L)}- S^{(L-1)}}{S^{(L-1)}} \right|.
\end{equation}
This yields five-loop values of the parameters $\alpha, b$ for each $u$.
In order to define 
an error bar, we apply the same method to the four-loop result of $B(u)$,
i.e.\ to $\Delta^{(3)}$ and $\Delta^{(2)}$.
The four-loop values of $\alpha, b$ together with the five-loop values
provide the ranges of the best values of $\alpha$ and $b$ as a result of
the combined four- and five-loop analysis. Then we 
define the error bar of the five-loop result
for $B(u)$ by the maximum and minimum of the resummed 
four- and five-loop values for $B(u)$ over the ranges of the best values of
$\alpha, b$. At the fixed point $u^\star$, Eqs.~(\ref{eq:u1})--(\ref{eq:u3}),
we find the ranges
\begin{eqnarray}
\label{eq:range1}
0.95 \le \alpha \le 1.08\,,\quad 5.7 \le b \le 7.75\quad \mbox{for } n=1\,,\\
\label{eq:range2}
0.94 \le \alpha \le 1.04\,,\quad 7.0 \le b \le 8.59\quad \mbox{for } n=2\,,\\
\label{eq:range3}
0.94 \le \alpha \le 1.02\,,\quad 8.18 \le b \le 9.76\quad \mbox{for } n=3\,.
\end{eqnarray}
%
%
The corresponding Borel resummed results for $B(u^\star)$ are
\begin{eqnarray}
\label{eq:B1}
B(u^\star) &=& 0.5024 \quad \pm 0.0011 \qquad \mbox{for }n=1\,, \\
\label{eq:B2}
B(u^\star) &=& 1.0053 \quad \pm 0.0022 \qquad \mbox{for }n=2\,, \\
\label{eq:B3}
B(u^\star) &=& 1.5080 \quad \pm 0.0034 \qquad \mbox{for }n=3\,. 
\end{eqnarray}
We have also determined the function $B(u)$ for 
$n = 2$ (superfluid $^4$He) in the range $0 \leq u \leq 0.04$ as shown
in Fig.\ \ref{figure:B_detail}.

Most remarkable is the smallness of the deviation of the resummed function
$B(u)$ for $n = 1,2,3$ from its two-loop approximation $n/2$.
This confirms previous conjectures \cite{SD2,KSD,HD} 
and justifies earlier analyses \cite{D2,D4,SD3}.

\section{Amplitude functions $F_\pm$ in three dimensions}
\label{sec:ampl_func}
\subsection{Definition of $F_\pm$}
\label{ssec:definition_F}
By means of the renormalized vertex functions in Eq.~(\ref{eq:C_ren}) we define
the dimensionless amplitude functions
\begin{equation}
\label{eq:F+-}
F_\pm(\mu\xi_\pm,u,d) = -4\mu^\e A_d^{-1}\Gamma_\pm^{(2,0)}(\xi_\pm,u,\mu,d)\,.
\end{equation}
They enter the critical behavior of the specific heat in three dimensions in 
the form of the functions
\begin{equation}
\label{eq:F(u)}
F_\pm(1,u,3) \equiv F_\pm(u)
\end{equation}
according to \cite{SD2}
\begin{equation}
\label{eq:crit_F}
\bare{C}^\pm = C_B + \frac{1}{4} a^2 \mu^{-1} A_3 K_\pm(u(l_\pm))
\exp \int_u^{u(l_\pm)} \frac{2\zeta_r(u')-1}{\beta_u(u',1)} \mbox{d}u'
\end{equation}
where
\begin{equation}
\label{eq:K(u)}
K_\pm(u) = F_\pm(u) - A(u,1)
\end{equation}
and $a=Z_r(u,1)^{-1} a_0$. In Eq.~(\ref{eq:crit_F}), $u(l)$ is the effective
coupling satisfying
\begin{equation}
\label{eq:u(l)}
l\, \frac{\mbox{d}u(l)}{\mbox{d}l} = \beta_u(u(l),1)
\end{equation}
with $u(1)=u$. The flow parameters $l_+$ and $l_-$ are chosen as
$l_+=(\mu\xi_+)^{-1}$ and $l_-=(\mu\xi_-)^{-1}$ above and below $T_\c$.

\subsection{Power series of $F_\pm$}
\label{ssec:series}
The amplitude functions are expandable in integer powers of $u$ \cite{SD2} and 
have the power series \cite{KSD}
\begin{equation}
\label{eq:F+(u)}
F_+(u) = \sum_{m=0}^\infty c^+_{Fm} u^m
\end{equation}
and \cite{HD}
\begin{equation}
  \label{eq:F-(u)}
F_-(u) = \frac{1}{u}\, \sum_{m=0}^\infty c^-_{Fm} u^m\,.
\end{equation}

We have determined $c_{Fm}^+$ up to five-loop order (i.e., up to $m=4$) for 
$n=1,2,3$ and $c_{Fm}^-$ up to five-loop order (i.e., up to $m=5$) for $n=1$
in two different ways. 

(i) The coefficients $c_{Fm}^+$ and $c_{Fm}^-$ can be calculated from 
Eqs.~(\ref{eq:F+-}),~(\ref{eq:C_ren}) in three dimensions according to
\begin{equation}
\label{eq:cF+-A}
F_{\pm}(u)= -16 \pi Z^2_r \xi_{\pm}^{-1} \bare{\Gamma}^{2,0}_{\pm}( \xi_{\pm},
         4 \pi \xi_{\pm}^{-1} Z_u Z_{\phi}^{-2}u, 3) + A(u, 1),
\end{equation}
where the Z-factors have the arguments $Z_i(u, 1)$. The perturbative expression
for $\bare{\Gamma}^{(2,0)}_+$ can be obtained for $n=1$, 2, 3 from 
\begin{equation}
\bare{\Gamma}^{(2,0)}_+(\xi_+, u_0, 3) = \frac{1}{4 u_0} Z_5^{-1}(\lambda)
\end{equation}
where the renormalization factor 
$Z_5(\lambda)$ and its relation to the specific heat have been presented
in numerical form
by Bervillier and Godr\`eche \cite{BG} and by Bagnuls and Bervillier 
\cite{BB,footnote1}, see also Ref.\ \cite{KSD}. 
For $d=3$ their renormalized coupling $\lambda$ is related to our $u_0$ 
via the renormalization factor $Z_3(\lambda)$ according to
$u_0 \xi_+= -2 \pi \lambda Z_3(\lambda)^{-1/2}$ as noted in 
Ref.~\cite{KSD}. 
The perturbative expression of $\bare{\Gamma}^{(2,0)}_-$ for $n=1$ can 
be determined according to 
\begin{eqnarray}
\bare{\Gamma}^{(2,0)}_-(\xi_-, u_0, 3)&=& 
\frac{\partial^2}{\partial r_0^{\prime 2}} \bare{\Gamma}_-(\xi_-, u_0, 3) 
\nonumber\\
&=& \left( \frac{\partial r_0^{\prime}}{\partial \xi_-} \right)^{-1} 
           \frac{\partial}{\partial \xi_-} \left[
    \left( \frac{\partial r_0^{\prime}}{\partial \xi_-} \right)^{-1} 
           \frac{\partial}{\partial \xi_-}
     \bare{\Gamma}_-(\xi_-, u_0, 3) \right],
\end{eqnarray}
where $r_0^{\prime}(\xi_-,u_0)$ is given by Eq.~(3.8) of Ref.~\cite{HD}. 
The Helmholtz free energy $\bare{\Gamma}_-(\xi_-, u_0, 3)$ 
is given in numerical form
by Eq.~(3.15) of Ref.\ \cite{HD} where our 
$\bare{\Gamma}_-(\xi_-, u_0, 3)$ is denoted by 
$\tilde{\Gamma}_{-0}(\xi_-, u_0)$.
Our numerical results for $c_{Fm}^{\pm}$ up to nine digits 
are presented in Table \ref{table:F}.

(ii) Alternatively the coefficients $c_{Fm}^{\pm}$ can be determined via the
relation
\begin{equation}
\label{eq:cF+-B}
8A_3^{-1} P_{\pm}(u) f_{\pm}^{(3,0)}(u) = (1-2 \zeta_r(u)) F_{\pm}(u)+ 4B(u) 
                                - \beta_u(u) \partial F_{\pm}(u)/ \partial u
\end{equation}
as done previously \cite{KSD,HD}. For the definition
of $P_{\pm}$ and $f_{\pm}^{(3,0)}$ and for a derivation of Eq.\ 
(\ref{eq:cF+-B}) we 
refer to Refs.\ \cite{SD1} and \cite{SD2}. In the present context we need the
contributions to $P_{\pm}$ and $f_{\pm}^{(3,0)}$ only up to $O(u^4)$
as given in Table 4 of Ref.~\cite{KSD} and Table 3 of Ref.~\cite{HD}
since $B(u)$ is known only up to $O(u^4)$ as well.
(We recall that the coefficients of $P_-$ are determined by those of $P_+$
according to $P_-(u)=-\frac{1}{2}\{1+2[1-P_+(u)]-\frac{3}{2}\zeta_r(u)\}$
\cite{SD2}.)
This calculation via Eq.~(\ref{eq:cF+-B}) 
yields coefficients $c_{Fm}^{\pm}$ that agree with those obtained via 
Eq.~(\ref{eq:cF+-A}) up to eight digits for $c_{Fm}^-$ and up to
seven digits for $c_{Fm}^+$.
The slight differences between the results of the calculations (i) and (ii)
are due to the fact that $Z_3(\lambda)$, $Z_5(\lambda)$ and $\bare{\Gamma}_-$
are available up to five-loop order only in numerical form.
We consider the calculation (i) via Eq.~(\ref{eq:cF+-A}) as 
slightly more reliable since 
fewer numerical operations are required than in the calculation (ii)
using Eq.~(\ref{eq:cF+-B}).    

Since here we have used the perturbative contributions of 
$B(u)$ up to five-loop order, the resulting 
higher-order
coefficients $c_{Fm}^{\pm}$ given in Table \ref{table:F}
differ from those determined previously 
(see Table~4 of Ref.~\cite{KSD} and Table~3 of Ref.~\cite{HD}) where the 
approximation $B(u)=n/2+O(u^2)$ was used. 
Only our low-order 
coefficients $c_{F0}^+$, $c_{F1}^+$, $c_{F0}^-$, $c_{F1}^-$ 
and $c_{F2}^-$ agree with the previous ones \cite{KSD,HD}.
The coefficients $c_{Fm}^+$ with $m>1$ differ considerably from the
previous ones whereas the coefficients $c_{Fm}^-$ with $m>2$ differ only
by 0.2\% $(m=3)$, 0.1\% $(m=4)$, and 2\% $(m=5)$.

Very recently the coefficients $c_{Fm}^+$ and $c_{Fm}^-$ have been determined
analytically for general $n$ up to three-loop order \cite{SLD}. The 
corresponding three-loop values $c_{F2}^+$ for $n=1$, 2, 3 as well as
$c_{F3}^-$ for $n=1$, 2, 3 are taken from Ref.~\cite{SLD} 
and are included in Table 
\ref{table:F} up to nine digits. No results for $c_{F3}^-$ were available
in the previous literature for $n>1$. The new information on $c_{F3}^-$
for $n>1$ enables us to perform the first Borel resummations for $F_-(u)$
and $F_-(u)-F_+(u)$ for $n=2$ and 3 in Subsection \ref{ssec:borel_3} below.

\subsection{Borel resummation of five-loop results for $n=1$}
\label{ssec:borel_5}
In order to study the effect of the new higher-order terms we have
performed Borel resummations of the series for $u F_-(u)$ and for
$u[F_-(u) - F_+(u)]$ at the fixed point $u^\star$ for the case $n = 1$.
The method employed is the same as for $B(u)$ in Sect. II. The 
parameter ranges turn out to be
\begin{equation}
\label{eq:alphab-}
1.6 \leq \alpha \leq 1.7\,, \quad 7.48 \leq b \leq 8.70
\end{equation}
for $u^\star F_-(u^\star)$, and
\begin{equation}
\label{eq:alphab+-}
1.4 \leq \alpha \leq 1.7\,, \quad 6.0 \leq b \leq 11.7
\end{equation}
for $u^\star [F_-(u^\star) - F_+(u^\star)]$.

We have found that our present method does not yield a reliable estimate
of the parameters $\alpha$ and $b$ for $F_+(u^\star)$ separately; 
this may be related to the
fact that, unlike $c^-_{Fm}$, the coefficients $c^+_{Fm}$ 
(see Table \ref{table:F})
do not have alternating signs for $m \leq 3$ (this alternation is predicted
for the asymptotic large-order behavior \cite{SD1,GZJ}). 
$F_+(u)$ will be further 
studied elsewhere. In the application to amplitude ratios given below
we shall not need $F_+(u^\star)$ separately.

The resummation results are
\begin{equation}
\label{eq:borelF-}
u^\star F_-(u^\star)=0.3687\pm0.0040
\end{equation}
and
\begin{equation}
\label{eq:borelF+-}
u^\star [ F_-(u^\star)-F_+(u^\star)]=0.4170\pm0.0036\,.
\end{equation}
The previous approximate resummation results \cite{KSD,HD} 
for $n = 1$ were $u^\star 
F_-(u^\star) = 0.3648$
and $u^\star [ F_-(u^\star)-F_+(u^\star)]=0.4170$ with an error bar
of about 1\%. 
Thus our resummation results differ from
the previous ones only by
about 1\% for $F_-$ and by less than 0.1\% for $F_--F_+$,
confirming previous conjectures \cite{KSD,HD}.
The parameter $d_F$ in the effective representation \cite{HD} $F_-(u)=(2u)^{-1}
-4(1+d_Fu)$ now becomes $d_F=-4.64$ (compared to $-4.04$ in Ref.\ \cite{HD}).
This leaves the solid line in Fig.~4 of Ref.~\cite{HD} essentially unchanged.

\subsection{Borel resummation of three-loop results for $n=2$ and $3$}
\label{ssec:borel_3}
While the previous two-loop result \cite{SD2} for $F_-(u)$ did not yet provide
sufficient information for a controlled resummation procedure for $n>1$ and
thus did not yet lead to an error estimate, the new three-loop coefficients
$c_{F3}^-$ have significantly improved the situation. On the basis of these
three-loop results we have performed Borel resummations of the series for
$uF_-(u)$ and for $u[F_-(u)-F_+(u)]$ for the cases $n=2$ and 3.
The method employed is the same as in Section \ref{sec:add_ren}. The parameter
ranges turn out to be
\begin{eqnarray}
\label{eq:alphab2-}
-1.70 &\leq& \alpha \leq 1.10\,, \quad 1.5 \leq b \leq 4.3\,, \quad n=2\,, \\
\label{eq:alphab3-}
-1.90 &\leq& \alpha \leq 1.15\,, \quad 1.5 \leq b \leq 4.7\,, \quad n=3
\end{eqnarray}
for $u^\star F_-(u^\star)$, and
\begin{eqnarray}
\label{eq:alphab2+-}
-0.80 &\leq& \alpha \leq 0.95\,, \quad 0.65 \leq b \leq 8.3\,, \quad n=2\,, \\
\label{eq:alphab3+-}
-0.30 &\leq& \alpha \leq 0.55\,, \quad 1.10 \leq b \leq 7.5\,, \quad n=3
\end{eqnarray}
for $u^\star [F_-(u^\star) - F_+(u^\star)]$.
The resummation results are
\begin{eqnarray}
\label{eq:borelF2-}
u^\star F_-(u^\star)&=&0.384\pm0.025\,, \quad n=2\,, \\
\label{eq:borelF3-}
u^\star F_-(u^\star)&=&0.387\pm0.026\,, \quad n=3\,,
\end{eqnarray}
and
\begin{eqnarray}
\label{eq:borelF2+-}
u^\star [ F_-(u^\star)-F_+(u^\star)]&=&0.461\pm0.019\,, \quad n=2\,, \\
\label{eq:borelF3+-}
u^\star [ F_-(u^\star)-F_+(u^\star)]&=&0.498\pm0.011\,, \quad n=3\,.
\end{eqnarray}

In order to substantiate the reliability of these resummations of three-loop
results we have repeated our resummation for $n=1$ but now on the basis of
three-loop results. In Fig.~4a we show the Borel resummation value for 
$uF_-(u)$ at $u=0.04$ for $n=1$ together with the corresponding error bar.
For comparison, the smaller error bar
resulting from the Borel summation on the basis of the five-loop results is
also shown. We see that the central value of the three-loop Borel resummation
is not far from the central 
value of the five-loop resummation (arrow in Fig.~4a). Analogous results are
shown in Fig.~5a for $u[F_-(u)-F_+(u)]$ at $u=0.04$ for $n=1$. 
These results provide confidence in
our resummation procedure based on three-loop results. 
In particular they demonstrate that Borel resummations of three-loop results
yield reliable results with error bars that significantly reduce the 
uncertainties of one- and two-loop calculations.
Since the $n$-dependence
beyond three-loop order is expected to be smooth and weak in the range
$1\leq n\leq3$ there is no reason to expect that the reliability of our
resummation procedure is significantly different for $n=2$ and 3. In Figs.~4b
and 5b the corresponding resummation results (with error bars) on the basis of
three-loop perturbation series are shown for
$n=2$ at the same value $u=0.04$ as for $n=1$ in Figs.~4a and 5a. 
They clearly demonstrate the significant
improvement over the previous situation at two-loop order \cite{SD2,BSD} where
no error bar could be determined in a convincing fashion.

\section{Applications}
\label{sec:applications}
We apply our results to the specific heat in the
asymptotic critical region where it can be represented as \cite{PHA,SD2}
\begin{equation}
\label{eq:C(t)}
\bare{C}^\pm=\frac{A^\pm}{\alpha} |t|^{-\alpha} 
\left ( 1 + a^\pm_c |t|^\Delta + \cdots  \right ) + B
\end{equation}
with the Wegner exponent \cite{W} $\Delta$. We consider the universal amplitude
ratios \cite{PHA} $A^+/A^-$ and $a^+_c/a^-_c$. 
The former can be expressed in terms
of $B^\star \equiv B(u^\star)$ and $F^\star_\pm \equiv  F_\pm (u^\star)$ in 
three dimensions
as \cite{SD2}
\begin{equation}
\label{eq:A+/A-}
\frac{A^+}{A^-} = 
\left( \frac{b_+}{b_-} \right)^\alpha 
\left[ 1- \alpha \frac{F^\star_- - F^\star_+}{4 \nu B^\star + \alpha F^\star_-}
\right]
\end{equation}
where
\begin{equation}
\label{eq:b+/b-}
\frac{b_+}{b_-} = \frac{2 \nu P^\star_+}{(3/2) - 2 \nu P^\star_+}
\end{equation}
with $P^\star_+\equiv P_+(u^\star)$. This expression for $A^+/A^-$ has several
independent sources of inaccuracies: (i) The values for the critical exponents
$\alpha$ and $\nu$, (ii) the values for $P_+^\star$, (iii) the values for
$B^\star$, (iv) the values for $F_-^\star-F_+^\star$ and $F_-^\star$.
In evaluating this expression for $A^+/A^-$ for $n=1$, 2, 3 we shall take (i)
the most reliable values for $\alpha$ and $\nu$ that are presently available,
(ii) the Borel resummed values for $P_+^\star$ based on five-loop results
for $n=1$, 2, 3, (iii) the Borel resummed values for $B^\star$ for $n=1$, 2, 3
as given by Eqs.~(\ref{eq:B1})--(\ref{eq:B3}), and (iv) the Borel resummed
values for $F_\pm^\star$ as given in Eqs.~(\ref{eq:borelF-}), 
(\ref{eq:borelF+-}), and (\ref{eq:borelF2-})--(\ref{eq:borelF3+-}).
For the critical exponents $\alpha$ and $\nu$ we take
\begin{eqnarray}
\label{eq:expo1}
\alpha&=&0.1070\pm0.0045\,, \quad \nu=0.6310\pm0.0015\,,\quad n=1\,, \\
\label{eq:expo2}
\alpha&=&-0.01285\pm0.00038\,, \quad \nu=0.67095\pm0.00013\,,\quad n=2\,, \\
\label{eq:expo3}
\alpha&=&-0.1150\pm0.0090\,, \quad \nu=0.7050\pm0.0030\,,\quad n=3\,,
\end{eqnarray}
according to Refs.~\cite{GZ}, \cite{LS} and \cite{GZ2}, respectively.
Borel summation values for $P_+^\star$ have been calculated previously
\cite{KSD}. These calculations, however, were based on five-loop coefficients
that were derived from the five-loop results of Refs.~\cite{CGLT} and
\cite{GLTC} which were corrected by Ref.~\cite{KS}. Taking into account the
latter corrections we have derived corrected five-loop coefficients $c_{P5}$
of the power series $P_+(u)=\sum_0^\infty c_{Pm} u^m$.
Our new values for $c_{P5}$ are (compare Table 4 of Ref.\cite{KSD})
\begin{eqnarray}
\label{eq:cP5_1}
c_{p5}&=&-1026631.56\,,\quad n=1\,,\\
\label{eq:cP5_2}
c_{p5}&=&-1763840.20\,,\quad n=2\,,\\
\label{eq:cP5_3}
c_{p5}&=&-2767879.03\,,\quad n=3\,.
\end{eqnarray}
For the corresponding Borel resummed values of $P_+^\star$ we have obtained
\begin{eqnarray}
\label{eq:borelP1}
P^\star_+&=&0.7568\pm0.0044\,,\quad n=1\,, \\
\label{eq:borelP2}
P^\star_+&=&0.7091\pm0.0045\,,\quad n=2\,, \\
\label{eq:borelP3}
P^\star_+&=&0.6709\pm0.0039\,,\quad n=3\,.
\end{eqnarray}
The values $P^\star_+$ obtained from Eq.~(3.2) and Table 1 of Ref.~\cite{KSD}
%
%
are by about 0.7\% smaller than the values in 
Eqs.~(\ref{eq:borelP1})--(\ref{eq:borelP3}).

If we use the central values of the critical exponents given by 
Eqs.~(\ref{eq:expo1})--(\ref{eq:expo3})
and collect the results of Eqs.~(\ref{eq:u1})--(\ref{eq:u3}),
(\ref{eq:B1})--(\ref{eq:B3}), (\ref{eq:borelF-}), (\ref{eq:borelF+-}),
(\ref{eq:borelF2-})--(\ref{eq:borelF3+-}), and
(\ref{eq:borelP1})--(\ref{eq:borelP3}) we arrive at the values for $A^+/A^-$
according to Eqs.~(\ref{eq:A+/A-}) and (\ref{eq:b+/b-}) as given in Table
\ref{table:ratios} for $n=1$, 2, 3.
For comparison this Table also contains the results of previous calculations
as well as experimental or numerical results. We see that there is good
agreement between the previous and our results for $n=1$ and $n=3$. For $n=2$,
however, our result is significantly more accurate and agrees well with the
high-precision experimental result for $^4$He near the superfluid transition
obtained from a recent experiment in space \cite{LS}.

The smallness of our error bar for $A^+/A^-$ for $n=2$ is due to the small
value of $\alpha$ for $n=2$ which, according to the structure of 
Eq.~(\ref{eq:A+/A-}), suppresses the error of $F_\pm^\star$.
Exploiting this structure, i.e., separating exponents from amplitude functions,
is a particular advantage of our $d=3$ formulation of field theory.
This structure
was not explicitly taken into account in the previous $\e$-expansion
analysis for $n=2$ \cite{B}.
The previous result
obtained from the $\e$ expansion up to $O(\e^2)$ \cite{B} does not agree with
the experimental result for $n=2$ 
within the error bars. The results 1.05 and 1.58 of
Ref.~\cite{SD2} for $n=2$ and $n=3$ are partly based on the one-loop form of
$F_-(u)$ for which no error bar was available in Ref.~\cite{SD2}.

The expression of $a^+_c/a^-_c$ is more complicated and depends on the 
derivatives
of the functions $F_-(u)$, $F_-(u)-F_+(u)$, $P_+(u)$, $B(u)$, and $\zeta_r(u)$
at the fixed point,
as specified in Eqs.~(4.24), (5.19) and (5.21)--(5.24) of Ref.\ \cite{SD2}.
We have performed Borel resummations for these quantities on the basis
of our new five-loop results for $n=1$.
For $a_c^+/a_c^-$ we obtain the value $1.0\,\pm\,0.1$. 
This is consistent with the previous result \cite{BBM}
$0.96\,\pm\,0.25$ for $n=1$. 
Since for $n>1$ only three-loop results are available (which would still
yield larger error bars for the derivatives of the various functions mentioned
above) we postpone corresponding calculations of $a_c^+/a_c^-$ for $n>1$
until four-loop results become available.

Finally we note that not only the fixed point values $B^\star$ and 
$F_\pm^\star$ but also the entire functions $B(u)$ and $F_\pm(u)$ are of
physical relevance. They are needed (i) for determining the effective 
renormalized coupling from experimental data of the specific heat \cite{D1,D2},
and (ii) for representing the specific heat in the entire nonasymptotic
critical region well away from $T_\c$ according to 
Eqs.~(\ref{eq:crit_F})--(\ref{eq:u(l)}). We also
note that $B^\star$ and $F^\star_\pm$ enter various other
important universal ratios, e.g.,
$R^T_\xi$ related to the superfluid density $\rho_s$ of $^4$He \cite{SD2}. 
For this ratio we refer to a recent paper \cite{SLD} where the three-loop
amplitude functions of the specific heat below $T_\c$ and of the order
parameter are derived for general $n$.

\vspace{5mm}
After completion of the present work we learned of the preprint
hep-ph/9710346 by B.\ Kastening, ``Five-Loop Vacuum
Energy Beta Function in $\phi^4$ Theory with O($n$)-Symmetric and Cubic
Interactions'' where the perturbative terms of a function equivalent
to $B(u)$ have been calculated up to five-loop order. 
These terms agree with ours in Eq.~(\ref{eq:B(u)}).
We have been informed by the author that this paper is accepted for
publication in Phys.~Rev.~D.
At the end of this work
it is asserted that in our paper we use the $\e$ expansion. Our work does not
use the $\e$ expansion to calculate the amplitude functions $F_\pm(u)$ and
the amplitude ratios $A^+/A^-$ and $a_c^+/a_c^-$. Only the fixed point value
$u^\star$ has been determined via a Borel resummation of the $\e$ expansion
series for $u^\star$.

\section*{Acknowledgements}
We gratefully acknowledge support by Deutsches Zentrum f\"ur Luft- und 
Raumfahrt (DLR, previously DARA) under grant 
number 50 WM 9669 as well as by NASA under contract number 960838.

\appendix
\newpage
\section{Derivation of the additive renormalization $A(\lowercase{u},\e)$}
\label{sec:vacuum}
The following expressions (\ref{eq:diff_1})--(\ref{eq:I121})
give the ultraviolet $d=4$ poles of the diagrams shown in 
Fig.~\ref{figure:vacuum} defined by $KR'(\partial^2 I^{(i)}/\partial r_0^2)$ 
according to the standard notations, see e.g.\ Ref.\ \cite{GLTC}.
Here $R'$ denotes the incomplete ultraviolet $R$-operation which subtracts
subdivergences, and $K$ denotes the operation of taking pole parts.
We use subscripts $(a.b)$ for the pole terms on the right-hand-sides of 
(\ref{eq:diff_1})--(\ref{eq:diff_8}) that are identical with the numbers
associated with the three- and four-loop diagrams of Ref.\ \cite{VKT} 
(the first number $a$ in the brackets indicates the number of loops and the
second number $b$ indicates the consecutive number of a diagram in Table 1
of Ref.\ \cite{VKT}). 
The subscripts on the right-hand sides of 
Eqs.~(\ref{eq:diff_9})--(\ref{eq:diff_18})
correspond to the numbers of the five-loop diagrams of Ref.~\cite{GLTC}.
We have multiplied the left-hand sides by factors
$(16\pi^2)^m$ in $m$-loop order because of the definition of the
bare four-point coupling $16\pi^2 g_0 / 24$ in Refs.~\cite{VKT,CGLT,GLTC,KS}.

The one- and two-loop expressions read
\begin{eqnarray}
\label{eq:diff_1}
16\pi^2 KR' \left( \frac{\partial^2}{\partial r_0^2}\; 
\left[ -\frac{1}{2} \int_p \ln (r_0+p^2) \right] \right) 
&=& \frac{1}{2}\, I_{(1.1)} \equiv J^{(1)} \,, \\
\label{eq:diff_2}
(16\pi^2)^2
KR' \left( \frac{\partial^2}{\partial r_0^2}\; I^{(2)}(r_0,\e) \right) 
&=& 2 I_{(2.2)} \equiv J^{(2)} \,,
\end{eqnarray}
the three-loop expressions read
\begin{eqnarray}
\label{eq:diff_3}
(16\pi^2)^3
KR' \left( \frac{\partial^2}{\partial r_0^2}\; I^{(3)}(r_0,\e) \right) 
&=& 2 I_{(3.2)} \equiv J^{(3)} \,, \\
\label{eq:diff_4}
(16\pi^2)^3
KR' \left( \frac{\partial^2}{\partial r_0^2}\; I^{(4)}(r_0,\e) \right) 
&=& 8 I_{(3.4)} +12 I_{(3.9)} \equiv J^{(4)} \,,
\end{eqnarray}
the four-loop expressions read
\begin{eqnarray}
\label{eq:diff_5}
(16\pi^2)^4
KR' \left( \frac{\partial^2}{\partial r_0^2}\; I^{(5)}(r_0,\e) \right) 
&=& 2 I_{(4.5)} \equiv J^{(5)} \,, \\
\label{eq:diff_6}
(16\pi^2)^4
KR' \left( \frac{\partial^2}{\partial r_0^2}\; I^{(6)}(r_0,\e) \right) 
&=& 0 \equiv J^{(6)} \,,\\
\label{eq:diff_7}
(16\pi^2)^4
KR' \left( \frac{\partial^2}{\partial r_0^2}\; I^{(7)}(r_0,\e) \right) 
&=& 4 I_{(4.9)} +6 I_{(4.11)} \equiv J^{(7)} \,,\\
\label{eq:diff_8}
(16\pi^2)^4
KR' \left( \frac{\partial^2}{\partial r_0^2}\; I^{(8)}(r_0,\e) \right) 
&=& 24 I_{(4.12)} +6 I_{(4.18)} +12 I_{(4.19)} \equiv J^{(8)} \,,
\end{eqnarray}
and the five-loop expressions read
\begin{eqnarray}
\label{eq:diff_9}
(16\pi^2)^5
KR' \left( \frac{\partial^2}{\partial r_0^2}\; I^{(9)}(r_0,\e) \right) 
&=& 2 I_{117} \equiv J^{(9)} \,, \\
\label{eq:diff_10}
(16\pi^2)^5
KR' \left( \frac{\partial^2}{\partial r_0^2}\; I^{(10)}(r_0,\e) \right) 
&=& 0 \equiv J^{(10)} \,,\\
\label{eq:diff_11}
(16\pi^2)^5
KR' \left( \frac{\partial^2}{\partial r_0^2}\; I^{(11)}(r_0,\e) \right) 
&=& 0 \equiv J^{(11)} \,,\\
\label{eq:diff_12}
(16\pi^2)^5
KR' \left( \frac{\partial^2}{\partial r_0^2}\; I^{(12)}(r_0,\e) \right) 
&=& 2 I_{121} \equiv J^{(12)} \,, \\
\label{eq:diff_13}
(16\pi^2)^5
KR' \left( \frac{\partial^2}{\partial r_0^2}\; I^{(13)}(r_0,\e) \right) 
&=& 4 I_7 +6 I_{120} \equiv J^{(13)} \,, \\
\label{eq:diff_14}
(16\pi^2)^5
KR' \left( \frac{\partial^2}{\partial r_0^2}\; I^{(14)}(r_0,\e) \right) 
&=& 2 I_8 \equiv J^{(14)} \,, \\
\label{eq:diff_15}
(16\pi^2)^5
KR' \left( \frac{\partial^2}{\partial r_0^2}\; I^{(15)}(r_0,\e) \right) 
&=& 2 I_{14} +12 I_{15} +12 I_{19} +24 I_{21} +4 I_{23} +18 I_{106} 
\equiv J^{(15)} \,,\\
\label{eq:diff_16}
(16\pi^2)^5
KR' \left( \frac{\partial^2}{\partial r_0^2}\; I^{(16)}(r_0,\e) \right) 
&=& 8 I_{79} +16 I_{88} +32 I_{95} +16 I_{100}\equiv J^{(16)}  \,, \\
\label{eq:diff_17}
(16\pi^2)^5
KR' \left( \frac{\partial^2}{\partial r_0^2}\; I^{(17)}(r_0,\e) \right) 
&=& 2 I_{81} +8 I_{84} +4 I_{99} \equiv J^{(17)} \,, \\
\label{eq:diff_18}
(16\pi^2)^5
KR' \left( \frac{\partial^2}{\partial r_0^2}\; I^{(18)}(r_0,\e) \right) 
&=& 
8 I_{32} +32 I_{47} +4 I_{73} + 8 I_{93} +8 I_{98} +4 I_{109} +8 I_{116} 
\equiv J^{(18)} \,. \nonumber\\
\end{eqnarray}
The pole terms up to four loops are \cite{VKT}
\begin{eqnarray}
\label{eq:(1.1)}
I_{(1.1)} &=& \frac{2}{\e} \,,\\
\label{eq:(2.2)}
I_{(2.2)} &=& -\frac{4}{\e^2} \,,\\
\label{eq:(3.2)}
I_{(3.2)} &=& \frac{8}{\e^3} \,,\\
\label{eq:(3.4)}
I_{(3.4)} &=& \frac{2}{3\e^2} -\frac{3}{4\e} \,,\\
\label{eq:(3.9)}
I_{(3.9)} &=& \frac{8}{3\e^3} -\frac{8}{3\e^2} +\frac{2}{3\e} \,,\\
\label{eq:(4.5)}
I_{(4.5)} &=& -\frac{16}{\e^4} \,,\\
\label{eq:(4.9)}
I_{(4.9)} &=& -\frac{4}{3\e^3} +\frac{3}{2\e^2} \,,\\
\label{eq:(4.11)}
I_{(4.11)} &=& -\frac{16}{3\e^4} +\frac{16}{3\e^3} -\frac{4}{3\e^2} \,,\\
\label{eq:(4.12)}
I_{(4.12)} &=& -\frac{4}{3\e^4} +\frac{10}{3\e^3} -\frac{13}{3\e^2} 
               +\frac{11-6\zeta(3)}{6\e} \,,\\
\label{eq:(4.18)}
I_{(4.18)} &=& -\frac{8}{3\e^4} +\frac{8}{3\e^3} +\frac{4}{3\e^2}
               +\frac{2\zeta(3)-2}{\e} \,,\\
\label{eq:(4.19)}
I_{(4.19)} &=& -\frac{2}{3\e^3} +\frac{1}{\e^2} -\frac{7}{12\e} \,,
\end{eqnarray}
and the five-loop pole terms are \cite{GLTC}
\begin{eqnarray}
\label{eq:I7}
I_7 &=& \frac{8}{3\e^4} -\frac{3}{\e^3} \,,\\
\label{eq:I8}
I_8 &=& \frac{8}{3\e^4} -\frac{3}{\e^3} \,,\\
\label{eq:I14}
I_{14} &=& \frac{4}{15\e^3} -\frac{3}{10\e^2} -\frac{5}{96\e} \,,\\
\label{eq:I15}
I_{15} &=& \frac{2}{15\e^3} -\frac{13}{20\e^2} +\frac{857}{960\e} \,,\\
\label{eq:I19}
I_{19} &=& \frac{8}{15\e^4} -\frac{7}{3\e^3} +\frac{25}{6\e^2} 
           -\frac{215}{96\e} \,,\\
\label{eq:I21}
I_{21} &=& \frac{16}{15\e^4} -\frac{7}{5\e^3} -\frac{11}{60\e^2}
           +\frac{157}{320\e} \,,\\
\label{eq:I23}
I_{23} &=& \frac{4}{15\e^3} -\frac{3}{10\e^2} -\frac{5}{96\e} \,,\\
\label{eq:I32}
I_{32} &=& \frac{48\zeta(3)}{5\e^3} -\frac{48\zeta(3)+24\zeta(4)}{5\e^2}
           +\frac{14\zeta(3)+9\zeta(4)-16\zeta(5)}{5\e} \,,\\
\label{eq:I47}
I_{47} &=& \frac{8}{15\e^5} -\frac{12}{5\e^4} +\frac{6}{\e^3}
         +\frac{18\zeta(3)-45}{5\e^2} +\frac{146-90\zeta(3)-9\zeta(4)}{30\e}
         \,,\\
\label{eq:I73}
I_{73} &=& \frac{16}{15\e^5} -\frac{8}{3\e^4} +\frac{28}{15\e^3}
         +\frac{6+4\zeta(3)}{5\e^2} - \frac{32-12\zeta(3)-18\zeta(4)}{30\e} 
         \,,\\
\label{eq:I79}
I_{79} &=& \frac{16}{5\e^5} -\frac{16}{5\e^4} -\frac{8}{5\e^3} 
           +\frac{4+4\zeta(3)}{5\e^2} + \frac{7+6\zeta(3)-12\zeta(4)}{5\e} 
           \,,\\
\label{eq:I81}
I_{81} &=& \frac{16}{3\e^5} -\frac{16}{3\e^4} -\frac{8}{3\e^3}
           +\frac{4-4\zeta(3)}{\e^2} \,,\\
\label{eq:I84}
I_{84} &=& \frac{8}{3\e^5} -\frac{20}{3\e^4} +\frac{26}{3\e^3}
           -\frac{44-24\zeta(3)}{12\e^2} \,,\\
\label{eq:I88}
I_{88} &=& \frac{16}{15\e^5} -\frac{8}{3\e^4} +\frac{28}{15\e^3} 
           +\frac{6-12\zeta(3)}{5\e^2} -\frac{16-18\zeta(4)}{15\e} \,,\\
\label{eq:I93}
I_{93} &=& \frac{8}{5\e^5} -\frac{52}{15\e^4} +\frac{34}{15\e^3}
           +\frac{116-24\zeta(3)}{60\e^2} 
           -\frac{56-44\zeta(3)+6\zeta(4)}{20\e} \,,\\
\label{eq:I95}
I_{95} &=& \frac{16}{15\e^5} -\frac{8}{3\e^4} +\frac{28}{15\e^3}
           +\frac{6-12\zeta(3)}{5\e^2} -\frac{16-18\zeta(4)}{15\e} \,,\\
\label{eq:I98}
I_{98} &=& \frac{4}{15\e^4} -\frac{14}{15\e^3} +\frac{19}{15\e^2} 
           -\frac{386+768\zeta(3)}{960\e} \,,\\
\label{eq:I99}
I_{99} &=& \frac{4}{3\e^4} -\frac{2}{\e^3} +\frac{7}{6\e^2} \,,\\
\label{eq:I100}
I_{100} &=& \frac{4}{5\e^4} -\frac{6}{5\e^3} +\frac{1}{5\e^2} 
            +\frac{81-48\zeta(3)}{160\e} \,,\\
\label{eq:I106}
I_{106} &=& \frac{64}{15\e^5} -\frac{32}{5\e^4} +\frac{8}{5\e^3}
            +\frac{16}{15\e^2} -\frac{2}{15\e} \,,\\
\label{eq:I109}
I_{109} &=& \frac{16}{15\e^5} -\frac{16}{5\e^4} +\frac{16}{5\e^3} 
            +\frac{52-108\zeta(3)}{15\e^2}
            -\frac{202-168\zeta(3)-18\zeta(4)}{30\e} \,,\\
\label{eq:I116}
I_{116} &=& \frac{8}{15\e^4} -\frac{4}{3\e^3} +\frac{32}{15\e^2}
            -\frac{250-96\zeta(3)}{120\e} \,,\\
\label{eq:I117}
I_{117} &=& \frac{32}{\e^5} \,,\\
\label{eq:I120}
I_{120} &=& \frac{32}{3\e^5} -\frac{32}{3\e^4} +\frac{8}{3\e^3} \,,\\
\label{eq:I121}
I_{121} &=& \frac{32}{3\e^5} -\frac{32}{3\e^4} +\frac{8}{3\e^3} \,.
\end{eqnarray}
In Eqs.\ (\ref{eq:I32}) and (\ref{eq:I116}) the corrections found in 
Ref.\ \cite{KS} have been taken into account. Note that in Eqs.\ 
(\ref{eq:(1.1)})--(\ref{eq:I121}) we have used $\e=4-d$ whereas in Refs.\
\cite{VKT,GLTC} $\e$ denotes $(4-d)/2$.

Eqs.~(\ref{eq:diff_1})--(\ref{eq:I121}) determine the additive renormalization
according to Eq.~(\ref{eq:C_ren}) as
\begin{eqnarray}
A(u,\e) &=& -2 \left[ \frac{n}{\e} - \frac{8n(n+2)}{\e^2}\: (-u/2) 
+ \sum_{i=3}^4 S^{(i)} G^{(i)} J^{(i)} (-u/2)^2 \right. \nonumber\\
\label{eq:A_final}
&&\mbox{}
+ \sum_{i=5}^8 S^{(i)} G^{(i)} J^{(i)} (-u/2)^3 \left.
+ \sum_{i=9}^{18} S^{(i)} G^{(i)} J^{(i)} (-u/2)^4 
                \right] \,.
\end{eqnarray}
The overall factor of 2 in Eq.~(\ref{eq:A_final})
arises from the $d=4$ value of the factor $(A_d/4)^{-1}/16\pi^2$
which is needed to obtain $A(u,\e)$ from Eqs.~(\ref{eq:diff_1})--(\ref{eq:I121})
according to Eq.~(\ref{eq:C_ren}).
The renormalized coupling $u$ 
enters Eq.~(\ref{eq:A_final}) in the form $u/2$; the factor $1/2$
takes into account that, near $d=4$, $u_0=A_d^{-1}u+O(u^2)=8\pi^2u+O(u^2)
=\frac{1}{2}[16\pi^2u+O(u^2)]$ [see Eq.~(\ref{eq:renorm})].

\newpage
\section{$Z$ factors}
\label{sec:factors}
In deriving the coefficients of the perturbation series of the 
quantities $F_\pm(u)$, $P_\pm(u)$, and $f^{(3,0)}_\pm(u)$
we needed the $Z$-factors $Z_r$, $Z_\phi$, and
$Z_u$ calculated previously \cite{CGLT,GLTC,KS} up to five-loop order. 
Since their explicit form is not available in the literature we present them
here explicitly. They read
\begin{eqnarray}
Z_r(u,\epsilon) &=& 1 + \sum^5_{m=1} Z_r^{(m)}(\epsilon) u^m + O(u^6)\,, \\
Z_u(u,\epsilon) &=& 1 + \sum^5_{m=1} Z_u^{(m)}(\epsilon) u^m + O(u^6)\,, \\
Z_\phi(u,\epsilon) &=& 1 + \sum^5_{m=1} Z_\phi^{(m)}(\epsilon) u^m + O(u^6)\,,
\end{eqnarray}
with the following coefficients in $m$-loop order:\\
Coefficients of $Z_r(u,\e)$:
\begin{eqnarray}
\label{eq:Zr1}
Z_r^{(1)}(\e) &=& \frac{4(n+2)}{\e} \,, \\
\label{eq:Zr2}
Z_r^{(2)}(\e) &=& 4(n+2) \left[ \frac{4(n+5)}{\e^2} -\frac{5}{\e} \right]
                  ,\\
\label{eq:Zr3}
Z_r^{(3)}(\e) &=& \frac{16}{3}\,(n+2) \left[
                  \frac {15n+111}{\e} + \frac {-278-61n}{\e^2}
                  + \frac {12n^2 +132n +360}{\e^3} \right] ,\\
\label{eq:Zr4}
Z_r^{(4)}(\e) &=& -\frac{2}{3}\,(n+2) \left[
                   \frac {288\zeta(4)(5n+22) +48\zeta(3)(3n^2 +10n +68) 
                   + 31060 -n^2 +7578n}{\e} \right. \nonumber\\
        && \mbox{} - \frac{1152\zeta(3)(22 +5n) +1236n^2 +23580n 
                   +74616}{\e^2} + \frac{16(245n^2 +2498n +6284)}{\e^3} 
                   \nonumber\\
        && \mbox{} \left. -192\,\frac {(n+5) (2n+13) (n+6)}{\e^4} \right] ,\\
\label{eq:Zr5}
Z_r^{(5)}(\e) &=& \frac{4}{15}(n+2) \Bigg[
                  \Big( 9600\zeta(6) (55n + 2n^2 +186) 
                   +768\zeta(5) (-5n^2 +72 +14n) \nonumber\\
       && \mbox{} +288\zeta(4)(29{n}^{2} + 2668 - 3n^3 +816n) 
                  +768\zeta(3)^2(-2n^2 -145n -582) \nonumber\\
       && \mbox{} +48 \zeta(3)(8208n +17n^3 +940n^2 + 31848) 
                  + 21n^3 +45254n^2 +1077120n \nonumber\\
       && \mbox{} + 3166528 \Big) \frac{1}{\e}
                  - \Big( 30720\zeta(5)( 2n^2 +186 +55n)
                  + 576\zeta(4)(5n + 22)(n - 22) \nonumber\\
       && \mbox{} + 96 \zeta(3)( 27n^3 +1224n^2 +14456n +45448) 
                  - 98n^3 + 277280n^2 +3073376n \nonumber\\
       && \mbox{} +7449712 \Big) \frac{1}{\e^2} 
                  +\Big( 2304\zeta(3)(13n + 74)(5n + 22) + 21576n^3 
                  + 685192n^2 \nonumber\\
       && \mbox{} + 5017312n + 10459360 \Big) \frac{1}{\e^3}
                  - \frac{32( 307976 + 31752n^2 + 172176n + 1933n^3}{\e^4} 
                  \nonumber \\
       && \mbox{} + \frac{384 (5n + 34)(n + 6)(n + 5)(2n + 13)}{\e^5} 
                  \Bigg] \,.
\end{eqnarray}
Coefficients of $Z_\phi(u,\e)$:
\begin{eqnarray}
\label{eq:Zphi1}
Z_\phi^{(1)} &=& 0 \,,\\
\label{eq:Zphi2}
Z_\phi^{(2)} &=& -\frac{4(n+2)}{\e} \,,\\
\label{eq:Zphi3}
Z_\phi^{(3)} &=& \frac{8}{3}\, (n+2)(n+8) \left[ \frac{1}{\e} 
               - \frac{4}{\e^2} \right] , \\
\label{eq:Zphi4}
Z_\phi^{(4)} &=&  2(n + 2) \left[ \frac{5(n^2 - 18n - 100)}{\e}
                + \frac{4(n^2 +234 +53n)}{\e^2} - \frac{16(n + 8)^2}{\e^3}
                \right]  ,\\
\label{eq:Zphi5}
Z_\phi^{(5)} &=& -\frac{8}{15}\,(n+ 2) \Bigg[
                \Big( -1152\zeta(4)(5n + 22) 
                + 48\zeta(3)( -6n^2 + 184 + n^3 + 64n)
                \nonumber \\
     && \mbox{} - 22752n - 39n^3 - 296n^2 - 77056 \Big) \frac{1}{\e} 
                \nonumber\\
     && \mbox{} + \frac{2304\zeta(3)(5n + 22) -60n^3 +135488 +42440n +1844n^2} 
                {\e^2} \nonumber \\
     && \mbox{} - \frac{16 (n + 8)(3n^2 + 269n + 1210)}{\e^3} 
                + \frac{192 (n + 8)^3}{\e^4} \Bigg] \,.
\end{eqnarray}
Coefficients of $Z_u(u,\e)$:
\begin{eqnarray}
\label{eq:Zu1}
Z_u^{(1)} &=& \frac{4(n+8)}{\e} \,,\\
\label{eq:Zu2}
Z_u^{(2)} &=& 16 \left[ \frac{(n+8)^2}{\e^2} -\, \frac{5n+22}{\e} \right] ,\\
\label{eq:Zu3} 
Z_u^{(3)} &=&  \frac{8}{3} \Bigg[
               \frac{96\zeta(3)(5n + 22) + 942n + 2992 + 35n^2}{\e}
               - \frac{16 (n + 8)(17n + 76)}{\e^2} \nonumber\\
   && \mbox{} + \frac{24 (n + 8)^3}{\e^3} \Bigg] \,,\\
\label{eq:Zu4}
Z_u^{(4)} &=& -\frac{16}{3} \Bigg[
              \Big( 480\zeta(5)(2n^2 + 55n + 186) 
              - 72 \zeta(4)(n + 8)(5n + 22) \nonumber\\
   && \mbox{} + 24 \zeta(3)(63n^2 + 764n + 2332) + 20624n + 1640n^2 - 5n^3 
              + 49912 \Big) \frac{1}{\e} \nonumber\\
   && \mbox{} - \frac{480 \zeta(3)(5n + 22)(n + 8) + 67424n + 153088 
              + 7736n^2 + 172n^3}{\e^2} \nonumber\\
   && \mbox{} + \frac{16 (55n + 248)(n + 8)^2}{\e^3}
              - \frac{ 48(n + 8)^4}{\e^4} \Bigg] \,, \\
\label{eq:Zu5}
Z_u^{(5)} &=& \frac{4}{15} \Bigg[ 
              \Big( 6912\zeta(7)(25774 + 9261n + 686n^2) 
              - 28800 \zeta(6)(n + 8)(2n^2 + 55n + 186) \nonumber\\
   && \mbox{} + 768\zeta(5)(165084 + 7466n^2 + 305n^3 + 66986n)
              - 288\zeta(4)(62656 + 4084n^2 + 28084n \nonumber\\
   && \mbox{} + 189n^3) + 2304\zeta(3)^2(3264 - 59n^2 - 6n^3 + 446n) 
              + 48\zeta(3)(1264n^3 - 13n^4 + 1312864 \nonumber\\
   && \mbox{} + 551032n + 67432n^2) + 20429248n + 2518864n^2 + 195n^4 
              + 40148480 + 39230n^3 \Big) \frac{1}{\e} \nonumber\\
   && \mbox{} - \Big( 99840 \zeta(5)(n + 8)(2n^2 + 55n + 186) 
              - 14976 \zeta(4)(5n + 22)(n + 8)^2 \nonumber\\
   && \mbox{} + 3456 \zeta(3)(91n^3 + 15436n + 34144 + 2196n^2) 
              + 63219712n - 800n^4 \nonumber\\
   && \mbox{} + 420800n^3 + 117768192 + 9811712n^2 \Big) \frac{1}{\e^2} 
              \nonumber\\
   && \mbox{} + \frac{66048\zeta(3)(5n + 22)(n + 8)^2
              + 32(n + 8)(733n^3 + 40186n^2 + 353392n + 803328)}{\e^3} 
              \nonumber\\
   && \mbox{} - \frac{512 (193n + 875)(n + 8)^3}{\e^4}
              + \frac{3840 (n + 8)^5}{\e^5} \Bigg]\,.
\end{eqnarray}

\begin{figure}
\caption{
Vacuum diagrams up to five-loop order determining the Helmholtz free energy
$\Gamma_0$ for $M_0=0$, $r_0>0$. Diagrams (6), (10) and (11) do not
contribute to $A(u,\e)$. The pole terms derived from the vacuum diagrams
are given in (\ref{eq:diff_1})--(\ref{eq:I121}) of App.~A.}
\label{figure:vacuum}
\end{figure}

\begin{figure}
\caption{
Partial sums $B_M(u) = \sum^M_{m=0} c_{{}_{Bm}} u^m$ of $B(u)$, Eq.\
(\ref{eq:power_B(u)}), 
as a function of $u$ for $n = 2$ from 
$M=1$ (two-loop order) to $M=4$ (five-loop order).
Also shown is the Borel resummed result
(solid line) which deviates from the two-loop result $B_1 = 1$ by
only 0.5 \% at the fixed point $u^\star = 0.0362$.}
\label{figure:B}
\end{figure}

\begin{figure}
\caption{
Borel resummation result for the function $B(u)$,
Eq.\ (\ref{eq:B(u)}), for $n = 2$ (solid line) obtained by interpolation between
the resummed values of $B(u)$ at $u_k=k\, u^\star/10$, $k=1,\ldots,10$
of the renormalized 
coupling $u$ in the range $0 < u \leq u^\star = 0.0362$, with error
bars. Also shown is the three-loop result $B_2(u) = 1 + 24 u^2$
(dashed line), compare Fig.~\ref{figure:B}. The two-loop result is $B_1 = 1$.
The Borel values $B(u_k)$ are 1.000233, 1.00073, 1.0013, 1.0019,
1.0026, 1.0032, 1.0037, 1.0043, 1.0048, 1.0053 for $k=1,\ldots,10$, 
respectively.}
\label{figure:B_detail}
\end{figure}

\begin{figure}
\caption{Amplitude function $F_-(u)$, multiplied by the renormalized coupling
$u$, in three dimensions as a function of $u$ in 1-, 2- and 3-loop order
(solid lines) for $n=1$ (a) and $n=2$ (b) and in 4- and 5-loop order for
$n=1$ (a) (dashed lines).
The dot with error bars at $u=0.04$ indicates the result of the Borel
resummation on the basis of the three-loop series for $uF_-(u)$.
The small error bar (a) indicates the Borel resummation result on the basis
of the five-loop series for $uF_-(u)$ at $u=0.04$ for $n=1$, the arrow
indicates the corresponding central value at $u=0.04$.
}
\label{figure:Fminus}
\end{figure}

\begin{figure}
\caption{Amplitude function $F_-(u)-F_+(u)$, 
multiplied by the renormalized coupling
$u$, in three dimensions as a function of $u$ in 1-, 2- and 3-loop order
(solid lines) for $n=1$ (a) and $n=2$ (b) and in 4- and 5-loop order for
$n=1$ (a) (dashed lines).
The dot with error bars at $u=0.04$ indicates the result of the Borel
resummation on the basis of the three-loop series for $u[F_-(u)-F_+(u)]$.
The small error bar (a) indicates the Borel resummation result on the basis
of the five-loop series for $u[F_-(u)-F_+(u)]$ 
at $u=0.04$ for $n=1$, the arrow indicates the corresponding central 
value at $u=0.04$.
}
\label{figure:Fplusminus}
\end{figure}

\newpage

\begin{table}
\caption{
Symmetry and group factors of the vacuum diagrams shown in 
Fig.~\protect\ref{figure:vacuum}. Diagrams (6), (10) and
(11) do not contribute to $A(u,\e)$.}
\begin{tabular}{cccc}
loop order & diagram $(i)$ & symmetry factor $S^{(i)}$ & group factor 
$G^{(i)}(n)$ \\
\tableline

1 loop & (1) & $1$ & $n$ \\
2 loops & (2) & $3 $ & $\frac{1}{3}(n^2+2n)$ \\
3 loops & (3) & $36 $ & $\frac{1}{9}(n^3+4n^2+4n)$ \\
       & (4) & $12 $ & $\frac{1}{3}(n^2+2n)$ \\
4 loops & (5) & $432 $ & $\frac{1}{27} (n^4+6n^3+12n^2+8n)$ \\
       & (7) & $576 $ & $\frac{1}{9}(n^3+4n^2+4n)$ \\
       & (8) & $288 $ & $\frac{1}{27} (n^3+10n^2+16n)$ \\
5 loops & (9) & $5184 $ & $\frac{1}{81} (n^5+8n^4+24n^3+32n^2+16n)$ \\
       & (12) & $10368 $ & $\frac{1}{27} (n^4+6n^3+12n^2+8n)$ \\
       & (13) & $6912 $ & $\frac{1}{27} (n^4+6n^3+12n^2+8n)$ \\
       & (14) & $6912 $ & $\frac{1}{27} (n^4+6n^3+12n^2+8n)$ \\
       & (15) & $2304 $ & $\frac{1}{9} (n^3+4n^2+4n)$ \\
       & (16) & $2592 $ & $\frac{1}{81} (n^4+8n^3+32n^2+40n)$ \\
       & (17) & $20736 $ & $\frac{1}{81} (n^4+12n^3+36n^2+32n)$ \\
       & (18) & $10368 $ & $\frac{1}{81} (5n^3+32n^2+44n)$
\end{tabular}
\label{table:symmetry}
\end{table}

\begin{table}
\caption{
Coefficients $f_i^{(m)}$ of the functions $\tilde{\beta}(u)$, 
$\zeta_r(u)$ and $\zeta_\phi(u)$ for $i=1, 2, 3$, respectively, and 
coefficients $c_{{}_{Bm}}$ of $B(u)$, compare Eqs.~(2.20) and (2.21), 
for $n=0,1,2,3$ up to five-loop order ($m=6$ for $\tilde{\beta}$, $m=5$ for
$\zeta_r$ and $\zeta_\phi$, $m=4$ for $B$).
For $f_i^{(m)}$ compare Table~1 of Ref.~\protect\cite{SD1}.} 
\begin{tabular}{cdddd}
 & $\tilde{\beta}_{u}$ & $\zeta_r$ & $\zeta_{\varphi}$ & $B$  \\
\tableline

$n=0$ & 0. & 8. & 0. & 0. \\
      &  32. &  -80.  & -16. & 0. \\
      &  -672. & 3552. & 128. & 0. \\
      & 43989.9534 & -223152.607 &  -8000. & 0. \\
      & -4166409.19 & 18836823.8 & 500639.112 & 0. \\
      &498653403.0 \\
\multicolumn{5}{c}{ } \\

$n=1$ &  0.  &  12. & 0. &  1/2 \\
      &  36. & -120. & -24.&  0. \\
      &  -816. &   6048. & 216. &  9. \\
      & 56245.8519 &-413813.942 &  -14040. & -761.422836\\
      & -5632017.54 & 37512804.7 &958294.321 & 44244.7100 \\
      & 708814936.0\\
\multicolumn{5}{c}{ } \\

$n=2$ &  0.  &  16.& 0. &  1.\\
      &  40. & -160. & -32. &  0.\\
      &  -960. & 9024.& 320. &  24.\\
      & 69029.7505 & -660870.017 &  -21120. & -2256.06766\\
      & -7268274.40 & 63662497.1 & 1566676.69 & 141294.329 \\
      & 956636505.0 \\
\multicolumn{5}{c}{ } \\

$n=3$ &  0. &  20. & 0. &  3/2 \\
      &  44. &  -200. & -40. &  0. \\
      &  -1104.  &  12480. & 440. &  45. \\
      & 82341.6490 & -967074.371 &  -29000.& -4653.13955  \\
      & -9075019.76 & 98265069.9 & 2333667.84 & 310944.846\\
      & 1243816220.0\\
\end{tabular}
\label{table1}
\end{table}
%
%

%

\begin{table}
\caption{Coefficients $c_{Fm}^\pm$ of $F_+(u)$ and $F_-(u)$ for $n=1$, 2, 3 
defined in Eqs.~(\ref{eq:F+(u)}) and (\ref{eq:F-(u)}), respectively. 
For $c_{Fm}^+$, $m$ refers to $u^m$ corresponding to $(m+1)$-loop order
whereas for $c_{Fm}^-$, 
$m$ refers to $u^{m-1}$ corresponding to $m$-loop order. 
The coefficients $c_{F2}^+$ and $c_{F3}^-$ (three-loop order)
are taken from Ref.~\protect\cite{SLD}}

\begin{tabular}{ccdd}
 & $m$ &  $c_{Fm}^+$ & $c_{Fm}^-$   \\
\tableline 
$n=1$  & 0 & -1.         & 1/2 \\
       & 1 & -6.         & -4. \\
       & 2 & -22.6976284 &  72. \\
       & 3 & -722.742498 & -5189.75474 \\
       & 4 &  34775.5861 &  433582.586 \\
       & 5 &             & -47754702.5 \\
\multicolumn{4}{c}{ } \\
$n=2$  & 0 & -2.         & 1/2 \\
       & 1 & -16.        & -4. \\
       & 2 & -92.5270090 & 64. \\
       & 3 & -2430.86460 & -5918.07320 \\
       & 4 & 102469.659 \\
\multicolumn{4}{c}{ } \\
$n=3$  & 0 & -3.         & 1/2 \\
       & 1 & -30.        & -4. \\
       & 2 & -233.488142 & 56. \\
       & 3 & -5742.02976 & -6607.95641 \\
       & 4 & 204463.777 \\

\end{tabular}
\label{table:F}
\end{table}

\begin{table}
\caption{Universal amplitude ratio $A^+/A^-$ of the specific heat, Eqs.\
(\ref{eq:C(t)})--(\ref{eq:b+/b-}), for $n=1$, 2, 3 in three dimensions.}
\begin{tabular}{c|d@{$\pm$}l|d@{$\pm$}l|d@{$\pm$}l|d@{$\pm$}l}
  & \multicolumn{4}{c|}{field theory\hspace*{1.5cm}} &
    \multicolumn{1}{c}{\raisebox{-2mm}{lattice series}}&& 
    \multicolumn{1}{c}{\raisebox{-5mm}{experiment}}\\
n & \multicolumn{2}{c|}{present work}&
    \multicolumn{1}{c}{previous work}&&
    \multicolumn{2}{c|}{\raisebox{3mm}{\hspace{-1cm}expansions}}&
    \multicolumn{1}{c}{}\\
\tableline
1 & 0.540& 0.011 
  & 0.524& 0.010\tablenotemark[1]
  & 0.523& 0.009\tablenotemark[3]
  &\multicolumn{2}{c}{0.56-0.63\tablenotemark[4], 0.53\tablenotemark[5]}\\
  &\multicolumn{1}{c}{} &
  & 0.541& 0.014\tablenotemark[2]& \multicolumn{1}{c}{}& \\ \tableline
2 & 1.056& 0.004 
  & 1.0294&0.0134\tablenotemark[1]
  & \multicolumn{1}{d}{1.08\tablenotemark[7]}&
  & 1.054 &0.001\tablenotemark[8]\\ 
  & \multicolumn{1}{c}{} &
  & \multicolumn{1}{d}{1.05\tablenotemark[6]} && \multicolumn{1}{c}{} &
  & 1.058& 0.004\tablenotemark[9] \\
  & \multicolumn{1}{c}{} & & \multicolumn{1}{c}{} & &
\multicolumn{1}{c}{}& 
  & 1.067 & 0.003\tablenotemark[10] \\
  & \multicolumn{1}{c}{} & & \multicolumn{1}{c}{} & &
\multicolumn{1}{c}{}& 
  & 1.088 & 0.007\tablenotemark[11] \\
\tableline
3 & 1.51 & 0.04  
  & 1.521& 0.022\tablenotemark[1] 
  & \multicolumn{1}{d}{1.52\tablenotemark[7]}&
  & \multicolumn{1}{d}{1.40\tablenotemark[12]} \\
  & \multicolumn{1}{c}{} &
  & \multicolumn{1}{d}{1.58\tablenotemark[6]} &
  & \multicolumn{1}{c}{} && \multicolumn{1}{c}{}\\
\end{tabular}
\tablenotetext[1]{Bervillier \protect\cite{B}}
\tablenotetext[2]{Bagnuls {\it et al.} \protect\cite{BBM}}
\tablenotetext[3]{Liu and Fisher \protect\cite{LF}}
\tablenotetext[4]{Beysens {\it et al.} \protect\cite{BBC}}
\tablenotetext[5]{Kumar {\it et al.} \protect\cite{KKG}}
\tablenotetext[6]{Schloms and Dohm \protect\cite{SD2}}
\tablenotetext[7]{Hohenberg {\it et al.} \protect\cite{HAHS}}
\tablenotetext[8]{Lipa {\it et al.} \protect\cite{LS}}
\tablenotetext[9]{Lipa and Chui \protect\cite{LC}}
\tablenotetext[10]{Singsaas and Ahlers \protect\cite{SA}}
\tablenotetext[11]{Takada and Watanabe \protect\cite{TW}}
\tablenotetext[12]{Kornblit and Ahlers \protect\cite{KA}}
\label{table:ratios}
\end{table}


\begin{thebibliography}{99}
\bibitem{footnote1}
In the headline of Table I of Ref.\ \cite{BB}, 
$Z_5^{-1}/\lambda$ has to be replaced by $Z_5^{-1}$.
\end{thebibliography}
\end{document}